\documentclass[onecolumn,preprintnumbers,amsmath,amssymb]{revtex4}
\usepackage{epsfig,graphicx}
\usepackage{dcolumn}
\usepackage{bm}
\begin{document}
\draft
\title{Computer Modeling of Electronic Properties
of Scroll-like V$_2$O$_5$-based Nanotubes}

\vspace{1cm}
\author{A.N. Enyashin, V.V. Ivanovskaya*, Yu.N. Makurin and A.L. Ivanovskii
\footnote[1]{For correspondence: viktoria@ihim.uran.ru}}
\address{Institute of Solid State Chemistry,\\ Ural Branch of the Russian Academy of
Sciences,\\Ekaterinburg, 620219, Russia}

\begin{abstract}

Atomic models of quasi-one-dimensional (1{\it D}) vanadium oxide
nanostructures - nanotubes of various morphology (cylinder or
scroll-like) formed by rolling (010) single layers of V$_2$O$_5$
are constructed and their electronic properties are studied using
the tight-binding band theory. Compared to the cylindrical {\it
zigzag} ({\it n},0) and {\it armchair} ({\it n},{\it n})-like
nanotubes, which are uniformly semi-conducting with the bang gap
of about 2.5-2.7 eV, the band gap of the scroll-like tubes trends
significantly to vanish (up to about 0.1 eV) depending on the
atomic configurations of the tubes and inter-wall distances.

\end{abstract}
\maketitle

% \begin{multicols}{2}

Nanostructured transition metal oxides present a unique class of
new materials with interesting electronic, chemical, and
catalytic properties. Recent advances in synthesis provide
appropriative methods for preparing vanadium oxide nanotubes (NT)
(review \cite{1}). For instance, hollow V$_2$O$_5$ nanotubes have
been produced using carbon NTs as templates \cite{2}. A wide
family of vanadium oxide (VO$_x$) nanotubes has been discovered
recently also as the main product of a sol-gel reaction that is
followed by hydrothermal treatment \cite{3,4}.

VO$_x$ NTs have attracted wide-spread attention due to their
potential application in catalysis \cite{2}, as nanoscale
materials for lithium batteries \cite{5,6} etc. Current studies
have been focused on the development of low-cost methods for
fabricating VO$_x$ nanotubular materials \cite{7,8}, and also on
investigation of their properties. Recently, Ivanovskaya et al
\cite{9} reported the first theoretical study of the electronic
structure of cylindrical single-walled vanadium pentoxide-based
NTs. It was established that both {\it zigzag}- and {\it
armchair}-like tubes are semiconducting. {\it Zigzag} nanotubes
were found to be more stable. V-O covalent bonds are the
strongest interactions, whereas V-V bonds are much weaker.

However most of the VO$_x$ NTs synthesized by template-directed
routes are multi-walled, the layer structure inside the tube
walls is frequently disordered, and several types of defects
appear. Moreover, as has been established by transmission
electron microscopy (TEM) cross-sectional images, the bent VO$_x$
layers inside the tube walls are preferentially scrolls rather
than concentric cylinders, and the wall structure of the tubes
contains organic molecules (amine, diamine etc.) as
structure-directing templates \cite{10,11,12}. Organic template
molecules are embedded between VO$_x$ layers, and the inter-layer
distances are proportional to the length of the molecules. In
particular, a unique type of VO$_x$ tubes with alternating
inter-layer distances is prepared by a template-assisted route
\cite{12}.

In the current paper, we performed atomic simulation of the
scroll-like morphology of nanotubes constructed from the
V$_2$O$_5$ (010) single plane. Their electronic band properties
were calculated and compared with cylindrical tubes in {\it
armchair}- and {\it zigzag}-like forms.

{\bf Atomic models.} We have considered the following structures
as models. The orthorhombic crystal V$_2$O$_5$ (space group {\it
Pmmn}) is built up by stacking 2{\it D}-like layers along (010)
and is composed by distorted VO$_5$ pyramids. There are three
structurally non-equivalent types of oxygen centers in V$_2$O$_5$:
single coordinated vanadyl oxygen O(1), double coordinated O(2),
and bridging oxygen O(3) triply coordinated to three vanadium
atoms \cite{13}, see Fig. 1. The structure of the (010)
single-layer slab is formed by VO$_5$ units sharing edges and
corners. We have analyzed the electronic properties of the
following 1{\it D} nanostructures, which can be constructed from
the mentioned above (010) slab: (I) planar atomic strips; (II)
cylindrical and (III) scroll-like NTs. The resulting NTs may be
described as
"fifty-wall"(O$^i$$^n$-V$^i$$^n$-O$^c$-V$^o$$^u$$^t$-O$^o$$^u$$^t$)
tubular structures. Here the "outer" and "inner" single-atomic
walls (O$^o$$^u$$^t$, O$^i$$^n$) are composed by vanadyl oxygen
atoms, and the "central" wall (O$^c$) is formed by O(2), O(3)
atoms. For the cylindrical vanadium pentoxide tubes, three groups
of structures can be constructed: non-chiral {\it armchair} ({\it
n},{\it n}), {\it zigzag} ({\it n},0)-, and chiral ({\it n},{\it
m})-like nanotubes (for detail see \cite{9}). Our calculations
were performed (using 365 atomic supercells) for (26,0) and
(26,26) NTs with inner diameters (D$^i$$^n$) 2.96 and 8.71 nm,
respectively, comparable to those observed experimentally (about
5-50 nm \cite{13}).

Models of V$_2$O$_5$ tubes with scroll-like morphology were
constructed based on the aforementioned cylindrical (26,0),
(26,26) NTs, which were "cut" along the tubular axis and rolled
up in scrolls. The calculations were performed for infinite-long
scrolls with two different inter-wall distances (minimal {\it
L}$_m$$_i$$_n$ = 0.45 and maximal {\it L}$_m$$_a$$_x$ = 1.50 nm)
chosen so that they are close to the inter-layer interval in the
bulk V$_2$O$_5$ (0.437 nm \cite{14}) and experimental inter-wall
distances of VO$_x$ tubes obtained with monoamine as a template
(1.5-1.7 nm \cite{3}), respectively. The cross-sectional
structure of the considered tube models is presented in Fig 1.

The tight binding band structure method within the extended
Huckel theory (EHT) approximation \cite{15} was employed to
establish the densities of states (DOS), Fermi energies (E$_F$),
and total band energies of the nanotubes (E$_t$$_o$$_t$). The
results obtained are given in Table 1 and in Figs. 2, 3.

{\bf Cylindrical V$_2$O$_5$ nanotubes.} The calculated DOS of the
cylindrical (26,0) and (26,26) V$_2$O$_5$ NTs (see Figs. 2, 3) are
similar to all other ({\it n},0) and ({\it n},{\it n})-like
vanadium pentoxide tubes (6$<${\it n}$<$13), respectively,
obtained earlier \cite{9}. For example, for the armchair-like
(26,26) tube the lowermost quasi-core band (not shown in Fig. 2)
consists mainly of O{\it 2s} states. The valence hybrid V{\it
3d}-O{\it 2p} band is fully occupied, and the DOS has three main
peaks (A, B, and C, Fig. 2) corresponding to the distribution of
electronic states of structurally non-equivalent atoms in the bulk
V$_2$O$_5$, see \cite{16}. The most intensive peak B below E$_F$
is formed by {\it 2p} states of vanadyl ions O(1), whereas {\it
2p} states of other oxygen atoms O(2,3) contribute to the whole
valence band with maxima near its edges (peaks A, C). According
to our estimates, the band gap (BG) of the (26,26) tube is about
2.7 eV. The lower part of the conduction band (peak D) is made up
predominantly of V{\it 3d} states.

The differences in (26,26) and (26,0)-like NTs electronic spectra
are connected with the DOS shape of the occupied V{\it 3d}-O{\it
2p} band and the lower energy part of the conduction band. From
Fig. 3 it is seen that for the zigzag-like (26,0)NT the DOS shape
is more close to that of the V$_2$O$_5$ (010) monolayer, in
particular, it does not contain the lower unoccupied separate
band (peak D, Fig.2). On the whole, it may be concluded that all
cylindrical tubes retain the semiconducting properties (Table 1),
and their electronic spectrum roughly resembles that of the
V$_2$O$_5$ planar strips considered as "precursors". The
electronic properties of scroll-like tubes differ essentially.

{\bf Scroll-like V$_2$O$_5$ nanotubes.} Fig. 2 displays the DOS of
scroll-like nanotubes (SL-NT) constructed on the basis of the
{\it armchair}-like (26,26) NT. The DOS of the SL-NT are seen to
have the following features: (I) new states are formed in the BG
region between the fully occupied valence V{\it 3d}-O{\it 2p} band
and the unoccupied V{\it 3d}-like conduction band; (II) the
occupation of these states leads to an increase in E$_F$; (III)
the lower unoccupied t$_2$$_g$-like band (separate peak D, Fig. 2)
is shifted upward the energy scale and merges with the lower edge
of the conduction band; and (IV) the DOS shape of the valence band
changes appreciably. The above effects are due both to the
appearance of new non-bonding states in the spectrum of SL-NTs
("dangling" bonds of V, O atoms located near the inner and outer
side of "sections lines" of the tubes) and partial restructuring
of the energy bands of SL-NTs resulting from different radii of
curvature of separate tube walls and inter-wall interactions. The
latter factor most distinctly affects the DOS of the SL-NT having
the minimum inter-wall distance ({\it L}$_m$$_i$$_n$ = 0.45 nm)
and the minimum diameter of the "inner" scroll, Figs. 2, 3.

As was shown in the calculation, the BG of the scrolls are much
smaller than those for cylindrical tubes of the corresponding
atomic dimensions and may vary considerably up to about 0.1 eV
(Table 1) depending on their structure. E$_t$$_o$$_t$ values
listed in Table 1 make it possible to perform semi-empirical
estimations of relative stability of 1{\it D} nanostructures
calculated using supercells of similar atomic dimensions (365
atoms per cell). So, for the family of nanostructures related to
the (26,0) tube, the cylindrical tubular form is more stable.
This can be qualitatively explained by the absence of "dangling"
bonds of V, O atoms. On the contrary, infinite-long scrolls are
more stable than the {\it armchair}-like (26,26) tube. It is
worth noting that according to our estimations for the various
infinite-long 1{\it D} nanostructures with the same size of cells
are shown, that (26,26)-like scrolls are more stable, see Table
1. Therefore it may be stated that scroll-like V$_2$O$_5$
nanostructures are generally more stable than "ideal" cylindrical
tubes, in accordance with recent experimental data.  It is
necessary to mention that we considered ideal "pure"
V$_2$O$_5$-based nanosructures. Certainly, the correct
description of energetic conditions of primary formation
experimentally observed VO$_x$ scrolls schould take into account
the "stabilizing" role of organic molecules  as templates.

In summary, atomic models of scroll-like V$_2$O$_5$ NTs have been
constructed and their electronic properties have been
investigated using the tight-binding band approach. We show that
their electronic spectrum features (including the BG) depend on
the scroll geometry and differ essentially from {\it zigzag}- and
{\it armchair}-like cylindrical nanotubes. Semi-empirical
estimations of E$_t$$_o$$_t$ suggest that infinite-long scrolls
should be more stable in comparison with ideal cylindrical
nanostructures.

Future computer simulations of VO$_x$ NTs should be aimed at
elaborating adequate theoretical models of actually produced
VO$_x$ nanotubular composites containing various organic template
molecules, and also at studying the effect of different tube wall
defects on the electronic properties of these materials.

{\bf Acknowledgment}

This work was supported by the Russian Foundation for Leading
Scientific Schools, grant HIII 829.2003.3

\newpage

\begin{table}
\begin{center}
\caption{Total energies (E$_t$$_o$$_t$, per V$_2$O$_5$ unit, eV),
Fermi energies (E$_F$, eV), and band gaps (BG, eV) of
infinite-long strips, single-walled cylindrical (26,0), (26,26)
vanadium pentoxide nanotubes and scrolls constructed from the
corresponding NTs with two different inter-wall distances ({\it
L}$_m$$_a$$_x$ = 1.50 and {\it L}$_m$$_i$$_n$ = 0.45 nm).}
\begin{center}
\begin{tabular}{|c|c|c|c|}
\hline  System &  -E$_t$$_o$$_t$/ V$_2$O$_5$  &  -E$_F$  &  BG  \\
\hline V$_2$O$_5$ monolayer & 787.896 & 14.30 & 2.70 \\
\hline V$_2$O$_5$ (26,0) strip & 787.812 &14.40&2.70\\
\hline V$_2$O$_5$ (26,0) scroll ({\it L}$_m$$_a$$_x$ =1.50 nm) &787.420&11.49&0.11\\
\hline V$_2$O$_5$ (26,0) scroll ({\it L}$_m$$_i$$_n$ = 0.45 nm) &787.402&12.97&1.60\\
\hline V$_2$O$_5$ (26,0) tube &787.896&14.30&2.91\\
\hline V$_2$O$_5$ (26,26) strip &787.244&14.47&2.76\\
\hline V$_2$O$_5$ (26,26) scroll ({\it L}$_m$$_a$$_x$ =1,50 nm) &788.428&12.56&1.76\\
\hline V$_2$O$_5$ (26,26) scroll ({\it L}$_m$$_i$$_n$ = 0.45 nm) &787.450&12.40&1.89\\
\hline V$_2$O$_5$ (26,26) tube &787.228&14.37&2.71\\
\hline
\end{tabular}
\end{center}

\end{center}
\end{table}

\newpage

\begin{figure}
\includegraphics[width=0.65\textwidth]{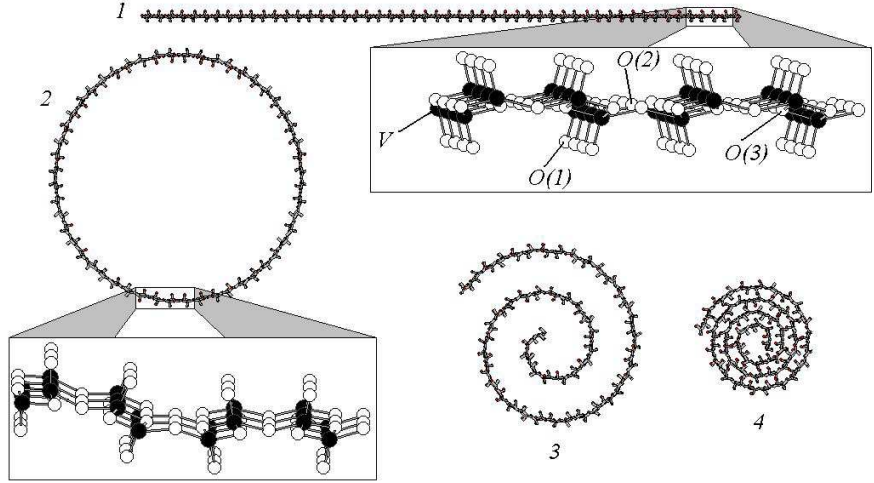}
\caption{Cross-sectional images of the structures of {\bf {\it
1}} - the (010) single layer strip of orthorhombic V$_2$O$_5$
(exhibited is a fragment of the strip, where positions of
non-equivalent oxygen atoms O(1)-O(3) are shown); {\bf {\it 2}} -
the cylindrical {\it armchair} (26,26)-tube (exhibited is a
fragment of the NT) and {\bf {\it 3, 4}} - scrolls constructed
from the {\it armchair} (26,26)-like V$_2$O$_5$ tube with two
different inter-wall distances ({\bf {\it 3}} - {\it
L}$_m$$_a$$_x$ = 1.50 and {\bf {\it 4}} - {\it L}$_m$$_i$$_n$ =
0.45 nm).}
\end{figure}

\begin{figure}
\includegraphics[width=0.65\textwidth]{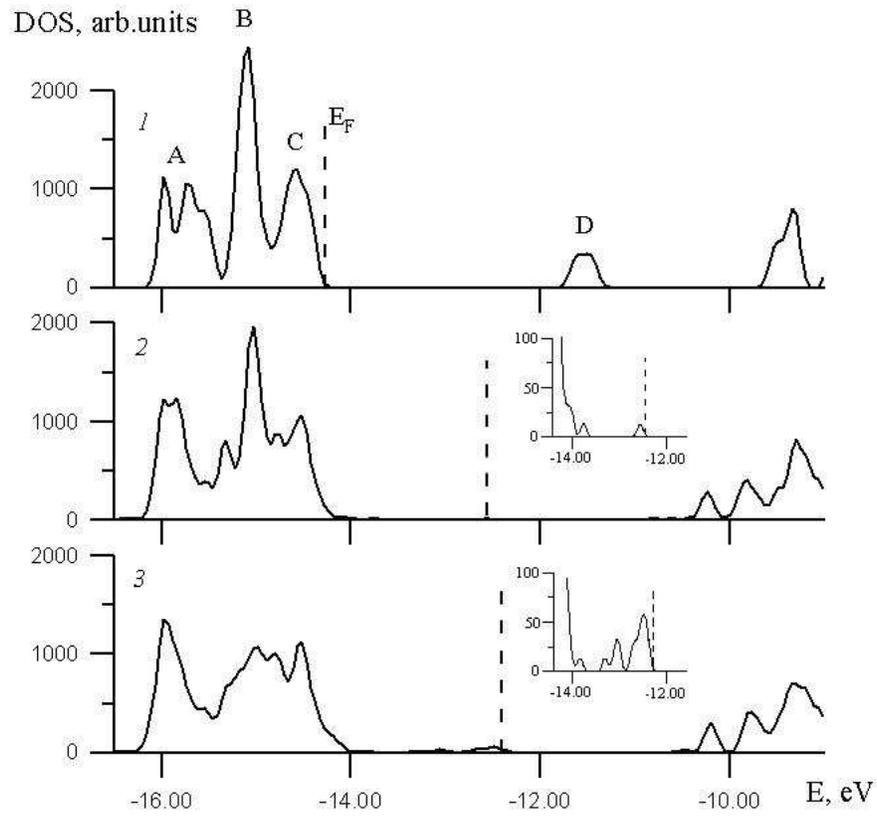}
\caption{Total DOS of the infinite-long {\bf {\it 1}} -
cylindrical {\it armchair} (26,26)-tube and {\bf {\it 2, 3}} -
scrolls constructed from the {\it armchair} (26,26)-like
V$_2$O$_5$ tube with two different inter-wall distances ({\bf
{\it 2}} - {\it L}$_m$$_a$$_x$ = 1.50 and {\bf {\it 3}} - {\it
L}$_m$$_i$$_n$ = 0.45 nm). }
\end{figure}

\begin{figure}
\includegraphics[width=0.65\textwidth]{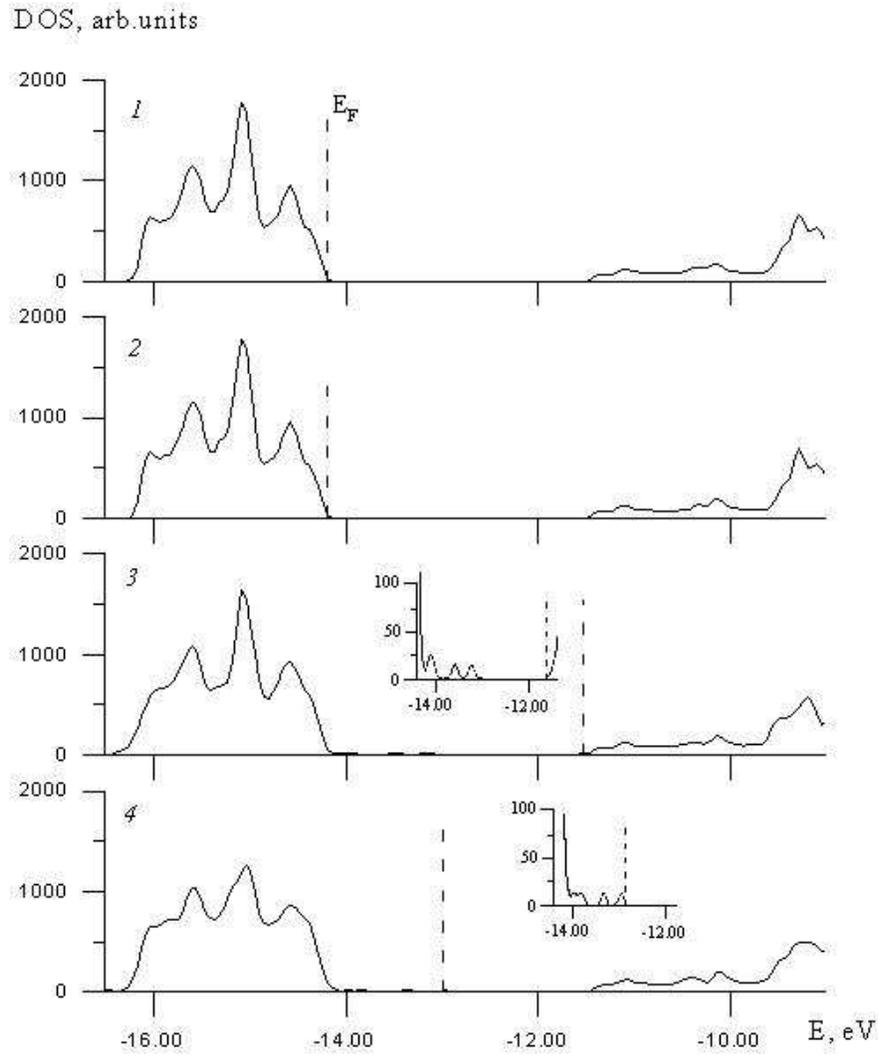}
\caption{Total DOS of the infinite-long {\bf {\it 1}} -
cylindrical {\it zigzag} (26,0)-tube and {\bf {\it 2}} - the
corresponding planar strip; {\bf {\it 3, 4}} - scrolls constructed
from the {\it zigzag} (26,0)-like V$_2$O$_5$ tube with two
different inter-wall distances ({\bf {\it 3}} - {\it
L}$_m$$_a$$_x$ = 1.50 and {\bf {\it 4}} - {\it L}$_m$$_i$$_n$ =
0.45 nm).}
\end{figure}

\end{document}